\documentclass[12pt, a4paper]{amsart}

\usepackage{amsfonts, amsmath, amssymb,latexsym}
\usepackage[english]{babel}
\usepackage[utf8]{inputenc}
\usepackage{url}
\usepackage[top=0.8in, left=1in, right=1in, bottom=1in]{geometry}

\usepackage{mathptmx}
\usepackage{times}

\def\R{{\Bbb R}}
\def\C{{\Bbb C}}
\def\H{{\Bbb H}}
\def\cl{{C}\!\ell}
\def\U{{\rm U}}
\def\T{{\rm T}}
\def\G{{\rm G}}
\def\L{{\rm L}}
\def\SU{{\rm SU}}
\def\diag{{\rm diag}}

\def\GL{{\rm GL}}
\def\SL{{\rm SL}}
\def\OO{{\rm O}}
\def\Sp{{\rm Sp}}
\def\sp{{\rm sp}}
\def\u{{\rm u}}
\def\so{{\rm so}}
\def\gl{{\rm gl}}
\def\Spin{{\rm Spin}}
\def\Mat{{\rm Mat}}
\def\tr{{\rm tr}}
 \newtheorem{thm}{Theorem}[section]

\begin{document}
\pagestyle{empty}
\title[On some Lie groups containing spin group in Clifford algebra]
{On some Lie groups containing spin group\\ in Clifford algebra}

\thanks{The reported study was funded by RFBR according to the research project No. 16-31-00347 mol\_a.}%

\maketitle

\vspace{-24pt}

\begin{center}
\author{{\bf D.~S.~Shirokov}\,$^a$ $^b$}%

\small

$^a$ Department of Mathematics, Faculty of Economic Sciences,\\ National Research University Higher School of Economics, Moscow, Russia,\\
dshirokov@hse.ru

\vspace{6pt}
$^b$ Kharkevich Institute for Information Transmission Problems,\\ Russian Academy of Sciences, Moscow, Russia,\\ shirokov@iitp.ru

\end{center}



\begin{abstract}
In this paper we consider some Lie groups in complexified Clifford algebras. Using relations between operations of conjugation in Clifford algebras and matrix operations we prove isomorphisms between these groups and classical matrix groups (symplectic, orthogonal, linear, unitary) in the cases of arbitrary dimension and arbitrary signature. Also we obtain isomorphisms of corresponding Lie algebras which are direct sums of subspaces of quaternion types. Spin group is a subgroup of all considered groups and it coincides with one of them in the cases $n\leq 5$. We present classical matrix Lie groups that contain spin group in the case of arbitrary dimension.
\end{abstract}

\thispagestyle{empty}

\section*{Introduction}

In this paper we prove isomorphisms between five Lie groups in complexified Clifford algebra and classical matrix groups in the case of arbitrary dimension and arbitrary signature. Also we obtain isomorphisms of corresponding Lie algebras. We further develop results of the paper \cite{first}.
In \cite{first} you can find statements only in the cases of fixed signatures when the corresponding real Clifford algebra has faithful and irreducible representations over $\R$ and $\R\oplus\R$. In the present paper we prove isomorphisms in all remaining cases, when real Clifford algebra has faithful and irreducible representations over $\C$, $\H$, and $\H\oplus\H$.

Lie groups which are considered in this paper can be useful in different questions of field theory. Note that the following three groups $\G^{2i1}_{p,q}$, $\G^{23}_{p,q}$, $\G^{2}_{p,q}$ are subgroups of pseudo-unitary group (see \cite{pseudou}, \cite{Snygg}, \cite{Dyab}). In \cite{Marchukeng} new symmetry of Dirac equation \cite{Dirac}, \cite{Dirac2} with respect to the pseudo-unitary group was considered.
Spin group $\Spin_{+}(p,q)$ is a subgroup of all five considered Lie groups. Moreover, group $\Spin_{+}(p,q)$ coincides with group $\G^{2}_{p,q}$ in the cases of dimensions $n\leq 5$. We discuss it in details in the Section \ref{SectionSpin}. Salingaros vee group (it consists of $\pm$ basis elements of real Clifford algebra) \cite{Sal1}, \cite{Sal2}, \cite{Sal3}, \cite{Helm2}, \cite{Abl2}, \cite{Abl3} is a subgroup of spin group and, so, also of five considered groups.

We mention a series of articles \cite{Abl1}, \cite{Abl2}, \cite{Abl3}, where many interesting facts about so-called transposition anti-involution are discussed. We consider such conjugation in real and complexified Clifford algebras and call it Hermitian conjugation in \cite{Marchuk:Shirokov}. In \cite{Marchuk:Shirokov} we were interested, mostly, in some particular problems related to applications in field theory; in \cite{Abl1}, \cite{Abl2}, \cite{Abl3} you can find more detailed description of corresponding algebraic structures. Note that operation of Hermitian conjugation of Clifford algebra elements is well-known, especially, in particular cases. For example, P. Dirac \cite{Dirac}, \cite{Dirac2} uses it in the case of signature $(p,q)=(1,3)$ in the theory of Dirac equation for electron. We should note that information about Hermitian conjugation in the Section 2 of the present paper is related to results of the paper \cite{Abl2}. In particular, one finds information about connection between Hermitian conjugation (or so-called transposition anti-involution) and matrix operations of corresponding matrix representations: in \cite{Abl2} for representations based on the fixed idempotent and the basis of corresponding left ideal, in the present paper (see formulas (\ref{relat})-(\ref{3rt})) for fixed matrix representations, in \cite{Marchuk:Shirokov} for complexified Clifford algebras and their representations based on the fixed idempotent and the basis of corresponding left ideal. In \cite{Abl2}, \cite{Abl3} some interesting facts about group $G_{p,q}^\epsilon=\{U\in\cl_{p,q}\, | \, U^\dagger U=e\}$ in real Clifford algebras were considered. In \cite{Marchuk:Shirokov} we consider analogue of this group $\{U\in\C\otimes\cl_{p,q}\, | \, U^\dagger U=e\}$ (so-called unitary group) in complexified Clifford algebras. The group $G_{p,q}^\epsilon$ is subgroup of this group. Note, that group $G^\epsilon_{p,q}$ coincides in particular cases with some of groups which are considered in the present paper: with the group $\G^{23}_{p,q}$ in the case of signature $(n,0$), with the group $\G^{12}_{p,q}$ in the case of signature $(0,n)$. Note, that some of groups which are considered in the present paper are related to automorphism groups of the scalar products on the spinor spaces (see \cite{Port}, \cite{Lounesto}, \cite{Abl3}), but we do not use this fact in the present paper. In \cite{Lounesto} one finds isomorphisms between groups $\G^{12}_{p,q}$, $\G^{23}_{p,q}$ and classical matrix Lie groups. In the present paper we also obtain these isomorphisms and also isomorphisms for groups $\G^{2i1}_{p,q}$, $\G^{2i3}_{p,q}$, $\G^{2}_{p,q}$ using other techniques based on relations between operations of conjugations in Clifford algebras and corresponding matrix operations. In particular, we generalized the notion of additional signature \cite{Pauli2} to the case of real Clifford algebras. We also study the corresponding Lie algebras with the use of techniques of quaternion types \cite{QuatAaca}, \cite{Quat2Aaca}, \cite{DAN} in Clifford algebras.

Let us consider the real Clifford algebra $\cl_{p,q}$ and complexified Clifford algebra $\C\otimes\cl_{p,q}$, $p+q=n$, $n\geq1$ \cite{Clifford}.
The construction of real and complexified Clifford algebras is discussed in details in \cite{Lounesto}, \cite{Chevalley}, \cite{Helm}.

Let us remind the basic notation. Let $e$ be the identity element and let $e^a$, $a=1,\ldots,n$ be generators\footnote{Note that $e^a$ is not exponent. We use notation with upper indices \cite{Benn:Tucker}.} of the Clifford algebra $\cl_{p,q}$,
$e^a e^b+e^b e^a=2\eta^{ab}e$, where $\eta=||\eta^{ab}||$ is the diagonal matrix with $+1$ appearing $p$ times on the diagonal and $-1$ appearing $q$ times on the diagonal. Elements
$e^{a_1\ldots a_k}=e^{a_1}\cdots e^{a_k}$, $a_1<\cdots<a_k$, $k=1,\ldots,n$, together with the identity element $e$, form a basis of Clifford algebra. Any Clifford algebra element $U\in\cl_{p,q}$ can be written in the form\footnote{We use Einstein's summation convention: there is a sum over index $a$.}
\begin{eqnarray}
U=ue+u_a e^a+\sum_{a_1<a_2}u_{a_1 a_2}e^{a_1 a_2}+\cdots+u_{1\ldots n}e^{1\ldots n},\label{uform}
\end{eqnarray}
where $u, u_a, u_{a_1 a_2}, \ldots, u_{1\ldots n}$ are real numbers. For arbitrary element $U\in\C\otimes\cl_{p,q}$ of complexified Clifford algebra we use the same notation (\ref{uform}), where $u, u_a, u_{a_1 a_2}, \ldots, u_{1\ldots n}$ are complex numbers.

 We denote by $\cl^k_{p,q}$ the vector spaces that span over the basis elements
$e^{a_1\ldots a_k}$. Elements of $\cl^k_{p,q}$ are said to be
elements of grade $k$. We have $\cl_{p,q}=\bigoplus_{k=0}^{n}\cl^k_{p,q}.$
Clifford algebra is a $Z_2$-graded algebra and it is represented as the direct sum of even and odd subspaces:
$$\cl_{p,q}=\cl^{(0)}_{p,q}\oplus\cl^{(1)}_{p,q},\quad \cl^{(i)}_{p,q}\cl^{(j)}_{p,q}\subseteq\cl^{(i+j)\,\rm{mod} 2}_{p,q},\quad \mbox{where}\quad \cl^{(i)}_{p,q}=\bigoplus_{k\equiv i\,\rm{mod}2}\cl^k_{p,q},\quad i, j=0, 1.$$

Let us consider the Clifford algebra $\cl_{p,q}$ as the vector space and represent it in the form of the direct sum of four subspaces of {\it quaternion types} 0, 1, 2 and 3 (see \cite{QuatAaca}, \cite{Quat2Aaca}, \cite{DAN}):
$$\cl_{p,q}=\overline{\textbf{0}}\oplus\overline{\textbf{1}}\oplus\overline{\textbf{2}}\oplus\overline{\textbf{3}},\qquad
\mbox{where}\quad \overline{\textbf{s}}=\bigoplus_{k\equiv s\,\rm{mod} 4}\cl^k_{p,q},\quad s=0, 1, 2, 3.$$
 We represent complexified Clifford algebra $\C\otimes\cl_{p,q}$ in the form of the direct sum of eight subspaces:
$\C\otimes\cl_{p,q}=\overline{\textbf{0}}\oplus\overline{\textbf{1}}\oplus\overline{\textbf{2}}\oplus\overline{\textbf{3}} \oplus i\overline{\textbf{0}}\oplus i\overline{\textbf{1}}\oplus i\overline{\textbf{2}}\oplus i\overline{\textbf{3}}.$

In \cite{first} we discussed a recurrent method of construction of matrix representation of real Clifford algebra in the cases of signatures $p-q\equiv0, 1, 2\mod 8$. In the Section \ref{sectmatr} of the present paper we generalize this method to the case of arbitrary signature.

In the Section \ref{sectherm} of the present paper we give some information about Hermitian conjugation in real and complexified Clifford algebras (see also \cite{Marchuk:Shirokov} and \cite{Abl2} for the case of real Clifford algebras). In \cite{Pauli2} we introduced the notion of additional signature $(k,l)$ of complexified Clifford algebra. In the Section \ref{sectadd} of the present paper  we generalize this notion to the case of real Clifford algebras.

In the Section \ref{sectmain} we prove isomorphisms between five Lie groups in Clifford algebra and classical matrix Lie groups (some of these isomorphisms are known, see above). We study corresponding Lie algebras. In the Section \ref{SectionSpin} of the present paper we discuss connection between groups $\Spin_{+}(p,q)$ and $\G^{2}_{p,q}$. In the Section \ref{sectconcl} we summarize results of \cite{first} and the present paper.

\section{Recurrent method of construction of matrix representations of real Clifford algebras in the case of arbitrary signature}\label{sectmatr}

In \cite{first} we discussed a recurrent method of construction of matrix representation only in the cases of signatures $p-q\equiv0, 1, 2\mod 8$ (the case of faithful and irreducible representations over $\R$ and $\R\oplus\R$). Now we generalize this method to the case of arbitrary signature. Clifford algebra has faithful and irreducible representations over $\R$, $\R\oplus\R$, $\C$, $\H$, $\H\oplus\H$ in different cases. However, items 1-4 are the same as in \cite{first}.

We have the following well-known isomorphisms $\cl_{p,q}\simeq \L_{p,q}$, where we denote by $\L_{p,q}$ the following matrix algebras
\begin{equation}
\L_{p,q}=\left\lbrace
\begin{array}{ll}
\Mat(2^{\frac{n}{2}},\R), & \parbox{.5\linewidth}{ if $p-q\equiv0; 2\!\!\mod 8$;}\\
\Mat(2^{\frac{n-1}{2}},\R)\oplus \Mat(2^{\frac{n-1}{2}},\R), & \parbox{.5\linewidth}{ if $p-q\equiv1\!\!\mod 8$;}\\
\Mat(2^{\frac{n-1}{2}},\C), & \parbox{.5\linewidth}{ if $p-q\equiv3; 7\!\!\mod 8$;}\\
\Mat(2^{\frac{n-2}{2}},\mathbb H), & \parbox{.5\linewidth}{ if $p-q\equiv4; 6\!\!\mod 8$;}\\
\Mat(2^{\frac{n-3}{2}},\mathbb H)\oplus \Mat(2^{\frac{n-3}{2}},\mathbb H), & \parbox{.5\linewidth}{ if $p-q\equiv5\!\!\mod 8$.}
\end{array}
\right.\nonumber
\end{equation}

We want to construct faithful and irreducible matrix representation $\beta: \cl_{p,q}\to \L_{p,q}$ of all real Clifford algebras with some additional properties related to symmetry and skew-symmetry of corresponding matrices (we discuss it in the next section). In some particular cases we construct $\beta$ in the following way:
\begin{itemize}
  \item In the case $\cl_{0,0}$: $e\to 1$.
  \item In the case $\cl_{0,1}$: $e\to 1$, $e^1\to i$.
  \item In the case $\cl_{1,0}$: $e\to {\bf1}=\diag(1,1)$, $e^1\to \diag(1,-1)$.
  \item In the case $\cl_{0,2}$: $e \to 1$, $e^1 \to i$, $e^2 \to j$.
  \item In the case $\cl_{0,3}$: $e\to {\bf1}=\diag(1,1)$, $e^1 \to \diag(i, -i)$, $e^2 \to \diag(j,-j)$, $e^3 \to \diag(k,-k)$.
\end{itemize}
For basis element $e^{a_1 \ldots a_k}$ we use the matrix that equals the product of matrices corresponding to generators $e^{a_1}, \ldots, e^{a_k}$. For identity element $e$, we always use identity matrix {\bf 1} of corresponding size.

Suppose that we have a faithful and irreducible matrix representation $\beta:\cl_{p,q}\to \L_{p,q}$
\begin{eqnarray}
e^a \to \beta^a,\quad a=1, \ldots, n,\label{matpred}
\end{eqnarray}
where $\beta^a=\beta(e^a)$, $a=1, 2, \ldots, n$. Now we want to construct the matrix representations of other Clifford algebras $\cl_{p+1, q+1}$, $\cl_{q+1, p-1}$, $\cl_{p-4, q+4}$ with generators $e^a$ ($a=1, 2, \ldots, n+2$ in the first case and $a=1, 2, \ldots, n$ in the last two cases) with the use of matrices $\beta^a$.  Using the following items 1-4 and Cartan periodicity of real Clifford algebras we obtain matrix representations of all real Clifford algebras. We call it recurrent method of construction of matrix representations\footnote{Another method of construction of matrix representations is based on the idempotent and basis of corresponding left ideal (see, for example, \cite{Abl1}, \cite{Abl2}, \cite{Abl3}, \cite{Marchuk:Shirokov}).}.

1.\quad Let us consider $\cl_{p+1, q+1}$. If $p-q\not\equiv 1\!\!\mod 4$, then for $p$ generators with squares $+1$ and $q$ generators with squares $-1$ we have
$$e^a\to \left( \begin{array}{ll}
 \beta^a & 0 \\
 0 & -\beta^a \end{array}\right),\qquad a=1, 2, \ldots, p, p+2, p+3, \ldots, p+q+1.$$
And for two remaining generators we have
$$
 e^{p+1}\to\left( \begin{array}{ll}
 0 & {\bf 1} \\
 {\bf 1} & 0 \end{array}\right),\quad
 e^{p+q+2}\to\left( \begin{array}{ll}
 0 & -{\bf 1} \\
 {\bf 1} & 0 \end{array}\right).
 $$

2. \quad If $p-q\equiv1\!\!\mod 4$, then matrices (\ref{matpred}) are block-diagonal and we have the following matrix representation of $\cl_{p+1, q+1}$. For $p+q$ generators $e^a$, $a=1, 2, \ldots, p, p+2, p+3, \ldots, p+q+1$ we have the same as in the previous item and for remaining two generators we have
 $$e^{p+1}\to
 \left( \begin{array}{ll}
 \beta^1\ldots\beta^n\Omega & 0 \\
 0 & -\beta^1\ldots\beta^n\Omega
  \end{array}\right),\quad
 e^{p+q+2}\to\left( \begin{array}{ll}
 \Omega & 0 \\
 0 & -\Omega
 \end{array}\right),
 $$
where
$$
\Omega =\left( \begin{array}{ll}
 0 & -{\bf 1} \\
 {\bf 1} & 0 \end{array}\right).
$$

3. \quad We construct matrix representation of $\cl_{q+1, p-1}$ using
$$e^1\to \beta^1,\qquad e^i\to \beta^i \beta^1,\quad i=2, \ldots, n.$$

4. \quad We construct matrix representation of $\cl_{p-4, q+4}$ using
$$e^i\to \beta^i \beta^1 \beta^2 \beta^3 \beta^4,\quad i=1, 2, 3, 4,\qquad e^j\to \beta^j,\quad j=5, \ldots, n.$$

In \cite{first} we gave some examples in the cases $p-q\equiv 0, 1, 2\mod 8$. Now let us give some examples in the cases of other signatures:

\begin{description}
  \item[$\cl_{1, 2}$]
 $$
  e^1 \to \left( \begin{array}{ll}
 0 & 1 \\
 1 & 0 \end{array}\right),\quad
 e^2 \to \left( \begin{array}{ll}
 i & 0 \\
 0 & -i \end{array}\right),\quad
 e^3 \to \left( \begin{array}{ll}
 0 & -1 \\
 1 & 0 \end{array}\right).
 $$
 \item[$\cl_{1, 3}$]
 $$
  e^1 \to \left( \begin{array}{ll}
 0 & 1 \\
 1 & 0 \end{array}\right),
 e^2 \to \left( \begin{array}{ll}
 i & 0 \\
 0 & -i \end{array}\right),
 e^3 \to \left( \begin{array}{ll}
 j & 0 \\
 0 & -j \end{array}\right),
 e^4 \to \left( \begin{array}{ll}
 0 & -1 \\
 1 & 0 \end{array}\right).
 $$
 \item[$\cl_{4, 0}$]
 $$
  e^1 \to \left( \begin{array}{ll}
 0 & 1 \\
 1 & 0 \end{array}\right),
 e^2 \to \left( \begin{array}{ll}
 0 & i \\
 -i & 0 \end{array}\right),
 e^3 \to \left( \begin{array}{ll}
 0 & j \\
 -j & 0 \end{array}\right),
 e^4 \to \left( \begin{array}{ll}
 -1 & 0 \\
 0 & 1 \end{array}\right).
 $$
 \item[$\cl_{0, 4}$]
 $$
  e^1 \to \left( \begin{array}{ll}
 k & 0 \\
 0 & -k \end{array}\right),
 e^2 \to \left( \begin{array}{ll}
 -j & 0 \\
 0 & -j \end{array}\right),
 e^3 \to \left( \begin{array}{ll}
 i & 0 \\
 0 & i \end{array}\right),
 e^4 \to \left( \begin{array}{ll}
 0 & k \\
 k & 0 \end{array}\right).
 $$
\end{description}

\section{Relation between operations of conjugation in Clifford algebra and matrix operations}\label{sectherm}

Consider the following well-known involutions in real $\cl_{p,q}$ and complexified Clifford algebra $\C\otimes\cl_{p,q}$:
$$
\hat{U}=U|_{e^a\to-e^a},\quad
\tilde{U}=U|_{e^{a_1\ldots a_r}\to e^{a_r}\ldots e^{a_1}},
$$
where $U$ has the form (\ref{uform}). The operation $U\to \hat{U}$ is called {\it grade involution} and $U\to \tilde{U}$ is called {\it reversion}.
Also we have operation of {\it complex conjugation}
\begin{eqnarray}
\bar U=\bar u e+\bar u_a e^a+\sum_{a_1<a_2}\bar u_{a_1 a_2}e^{a_1
a_2}+\sum_{a_1<a_2<a_3}\bar u_{a_1 a_2 a_3}e^{a_1 a_2
a_3}+\ldots,\nonumber
\end{eqnarray}
where we take complex conjugation of complex numbers $u_{a_1 \ldots a_k}$. Superposition of reversion and complex conjugation is {\em pseudo-Hermitian conjugation of Clifford algebra elements}\footnote{Pseudo-Hermitian conjugation of Clifford algebra elements is related to pseudo-unitary matrix groups as Hermitian conjugation is related to unitary groups, see \cite{Snygg}, \cite{Marchuk:Shirokov}.}
$$
U^\ddagger=\tilde{\bar{U}}.
$$
In the real Clifford algebra $\cl_{p,q}\subset \C\otimes\cl_{p,q}$ we have $U^\ddagger=\tilde{U}$, because $\bar U=U$.

Let us consider in complexified $\C\otimes\cl_{p,q}$ and real $\cl_{p,q}\subset\C\otimes\cl_{p,q}$ Clifford algebras the linear operation (involution) $\dagger: \C\otimes\cl_{p,q}\to\C\otimes\cl_{p,q}$ such that $(\lambda e^{a_1 \ldots a_k})^\dagger=\bar{\lambda} (e^{a_1 \ldots a_k})^{-1}$, $\lambda\in\C$. We call this operation {\it Hermitian conjugation of Clifford algebra elements}\footnote{Note that it is not Hermitian conjugation of matrix but it is related to this operation.}. We use notation $e_a=\eta_{ab}e^b=(e^a)^{-1}=(e^a)^\dagger$ and $e_{a_1\ldots a_k}=(e^{a_1 \ldots a_k})^{-1}=(e^{a_1 \ldots a_k})^\dagger$. This operation is well-known and many authors use it, for example, in different questions of field theory in the case of signature $(p,q)=(1,3)$. For more details, see \cite{Marchuk:Shirokov}, and for the case of real Clifford algebras see \cite{Abl1}, \cite{Abl2}, \cite{Abl3} (so-called transposition anti-involution in real Clifford algebras).

Note that we have the following relation between operation of Hermitian conjugation of Clifford algebra elements $\dagger$ and other operations in complexified Clifford algebra $\C\otimes\cl_{p,q}$ (see \cite{Marchuk:Shirokov})
\begin{eqnarray}
U^\dagger&=&e_{1\ldots p}U^\ddagger e^{1 \ldots p},\qquad \mbox{if $p$ is odd},\nonumber\\
U^\dagger&=&e_{1\ldots p}\hat{U}^{\ddagger} e^{1 \ldots p},\qquad \mbox{if $p$ is even},\label{sogldager2}\\
U^\dagger&=&e_{p+1\ldots n}U^\ddagger e^{p+1 \ldots n},\qquad \mbox{if $q$ is even},\nonumber\\
U^\dagger&=&e_{p+1\ldots n}\hat{U}^{\ddagger} e^{p+1 \ldots n},\qquad \mbox{if $q$ is odd}.\nonumber
\end{eqnarray}

We have the following well-known isomorphisms:
\begin{eqnarray}
\C\otimes\cl_{p,q}&\simeq& \Mat(2^{\frac{n}{2}}, \C),\qquad \mbox{if $n$ is even},\nonumber\\
\C\otimes\cl_{p,q}&\simeq& \Mat(2^{\frac{n-1}{2}}, \C)\oplus \Mat(2^{\frac{n-1}{2}}, \C),\qquad \mbox{if $n$ is odd}.\nonumber
\end{eqnarray}

Hermitian conjugation of Clifford algebra elements corresponds to Hermitian conjugation of matrix $\beta(U^\dagger)=(\beta(U))^\dagger$ for faithful and irreducible matrix representations over $\C$ and $\C\oplus\C$ of complexified Clifford algebra, based on the fixed idempotent and basis of corresponding left ideal, see \cite{Marchuk:Shirokov}.

It is not difficult to prove that for matrix representation $\beta$ of real Clifford algebra $\cl_{p,q}$ from the previous section we have
\begin{eqnarray}
\beta(U^\dagger)&=&(\beta(U))^T,\qquad p-q\equiv0, 1, 2\!\!\mod 8,\nonumber\\
\beta(U^\dagger)&=&(\beta(U)^\dagger,\qquad p-q\equiv3, 7\!\!\mod 8,\label{relat}\\
\beta(U^\dagger)&=&(\beta(U))^*,\qquad p-q\equiv4, 5, 6\!\!\mod 8,\nonumber
\end{eqnarray}
where $T$ is the operation of matrix transposition, $*$ is the operation of conjugate transpose of matrices of quaternions, $U^\dagger$ is the Hermitian conjugate of a Clifford algebra element $U$, $\beta(U)^\dagger$ is the Hermitian conjugate of the corresponding matrix.
Note, that in \cite{Abl2} analogous statement is proved for matrix representations based on the fixed idempotent and the basis of corresponding left ideal. In this paper we do not use idempotent to construct matrix representation. We use the recurrent method of construction of matrix representation (see previous section). So, we must verify statement for fixed matrices in some cases of small $n$ (see the beginning of the previous section) and then, using the method of mathematical induction, we must verify that, using items 1-4, we obtain matrix representations of other Clifford algebras with the same property (\ref{relat}). We omit the detailed proof.

In the cases $p-q\equiv0, 1, 2\!\!\mod 8$ for the real Clifford algebra $\cl_{p,q}\subset\C\otimes\cl_{p,q}$ we have
\begin{eqnarray}
U^T&=&e_{1 \ldots p} \tilde{U} e^{1 \ldots p},\qquad \mbox{if $p$ is odd},\nonumber\\
U^T&=&e_{1 \ldots p} \tilde{\hat{U}} e^{1 \ldots p},\qquad \mbox{if $p$ is even},\label{0rt}\\
U^T&=&e_{p+1 \ldots n} \tilde{U} e^{p+1 \ldots n},\qquad \mbox{if $q$ is even},\nonumber\\
U^T&=&e_{p+1 \ldots n} \tilde{\hat{U}} e^{p+1 \ldots n},\qquad \mbox{if $q$ is odd,}\nonumber
\end{eqnarray}
where $U^T=\beta^{-1}(\beta^T(U))=U^\dagger$.

In the cases $p-q\equiv3,7\!\!\mod 8$ for the real Clifford algebra $\cl_{p,q}$ we have
\begin{eqnarray}
U^\dagger&=&e_{1\ldots p}\tilde{U} e^{1 \ldots p},\qquad \mbox{if $p$ is odd},\nonumber\\
U^\dagger&=&e_{1\ldots p}\tilde{\hat{U}} e^{1 \ldots p},\qquad \mbox{if $p$ is even},\label{1rt}\\
U^\dagger&=&e_{p+1\ldots n}\tilde{U} e^{p+1 \ldots n},\qquad \mbox{if $q$ is even},\nonumber\\
U^\dagger&=&e_{p+1\ldots n}\tilde{\hat{U}} e^{p+1 \ldots n},\qquad \mbox{if $q$ is odd}.\nonumber
\end{eqnarray}

In the cases $p-q\equiv4, 5, 6\!\!\mod 8$ for the real Clifford algebra we have
\begin{eqnarray}
U^*&=&e_{1\ldots p}\tilde{U}e^{1 \ldots p},\qquad \mbox{if $p$ is odd},\nonumber\\
U^*&=&e_{1\ldots p}\tilde{\hat{U}} e^{1 \ldots p},\qquad \mbox{if $p$ is even},\label{3rt}\\
U^*&=&e_{p+1\ldots n}\tilde{U} e^{p+1 \ldots n},\qquad \mbox{if $q$ is even},\nonumber\\
U^*&=&e_{p+1\ldots n}\tilde{\hat{U}} e^{p+1 \ldots n},\qquad \mbox{if $q$ is odd},\nonumber
\end{eqnarray}
where $U^*=\beta^{-1}(\beta^*(U))=U^\dagger$. We use formulas (\ref{0rt}) - (\ref{3rt}) in the next sections.

\section{Additional signature of real Clifford algebra}\label{sectadd}

In \cite{Pauli2} we introduced the notion of additional signature $(k,l)$ of complexified Clifford algebra. Now we want to generalize this notion to the case of real Clifford algebras.

Suppose we have faithful and irreducible matrix representation $\beta$ over $\C$ or $\C\oplus\C$ of complexified Clifford algebra. We can always use such matrix representation that all matrices $\beta^a=\beta(e^a)$ are symmetric or skew-symmetric. Let $k$ be the number of symmetric matrices among $\{ \beta^a \} $ for matrix representation $\beta$, and $l$ be the number of skew-symmetric matrices among $\{ \beta^a \}$. Let $e^{b_1},\ldots, e^{b_k}$ denote the generators for which the matrices are symmetric. Analogously, we have $e^{c_1},\ldots, e^{c_l}$ for skew-symmetric matrices.

We use the notion of additional signature of Clifford algebra when we study relation between matrix representation and operations of conjugation.
In complexified Clifford algebra we have (see \cite{Pauli2})
\begin{eqnarray}
U^T&=&e_{b_1 \ldots b_k} \tilde{U} e^{b_1\ldots b_k},\qquad \mbox{$k$ is odd},\nonumber\\
U^T&=&e_{b_1 \ldots b_k} \tilde{\hat{U}} e^{b_1\ldots b_k},\qquad \mbox{$k$ is even},\label{sogltransp}\\
U^T&=&e_{c_1 \ldots c_l} \tilde{U}e^{c_1\ldots c_l},\qquad \mbox{$l$  is even},\nonumber\\
U^T&=&e_{c_1 \ldots c_l} \tilde{\hat{U}} e^{c_1\ldots c_l},\qquad \mbox{$l$ is odd.}\nonumber
\end{eqnarray}

Numbers $k$ and $l$ depend on the matrix representation $\beta$. But they can take only certain values despite dependence on the matrix representation.

In \cite{Pauli2} we proved that in complexified Clifford algebra we have only the following possible values of additional signature:
$$\begin{array}{ll}
\begin{tabular}{|c|c|}
\hline
$n\mod 8$ & $(k\mod 4, l\mod 4)$   \\ \hline \hline
$0$ & $(0,0),\,(1,3)$  \\ \hline
$1$ & $(1,0)$   \\ \hline
$2$ & $(1,1),\,(2,0)$  \\ \hline
$3$ & $(2,1)$   \\ \hline
\end{tabular} &
\begin{tabular}{|c|c|}
\hline
$n\mod 8$ & $(k\mod 4, l\mod 4)$   \\ \hline \hline
$4$ & $(3,1),\,(2,2)$   \\ \hline
$5$ & $(3,2)$   \\ \hline
$6$ & $(3,3),\,(0,2)$   \\ \hline
$7$ & $(0,3)$   \\ \hline
\end{tabular}
\end{array}
$$

Now we want to use the notion of additional signature in real Clifford algebra. Let us consider faithful and irreducible matrix representation over $\R$, $\R\oplus\R$, $\C$, $\H$, or $\H\oplus\H$ of real Clifford algebra. We can always use such matrix representation that all matrices $\beta^a=\beta(e^a)$ are symmetric or skew-symmetric (for example, we can use matrix representation $\beta:\cl_{p,q}\to \L_{p,q}$ from the Section \ref{sectmatr}). Let $k$ be the number of symmetric matrices among $\{ \beta^a \} $ for matrix representation $\beta$, and $l$ be the number of skew-symmetric matrices among $\{ \beta^a \}$. Let $e^{b_1},\ldots, e^{b_k}$ denote the generators for which the matrices are symmetric. Analogously, we have $e^{c_1},\ldots, e^{c_l}$ for skew-symmetric matrices.

Note, that in the cases $p-q\equiv0, 1, 2\!\!\mod 8$ for the real Clifford algebra $\cl_{p,q}$ the formulas (\ref{sogltransp}) are valid and they coincide with formulas (\ref{0rt}) (we have $p=k$ and $q=l$). In the cases $p-q\equiv3,7\!\!\mod 8$ for the real Clifford algebra we can use formulas (\ref{sogltransp}). The proof is similar to the proof for the complexified Clifford algebra. Actually, it is sufficient to prove these formulas for basis elements because of linearity of operation of conjugation. For example, if $k$ is odd, then
\begin{eqnarray}
(e^{a_1 \ldots a_m})^T&=&(e^{b_k})^{-1}\ldots (e^{b_1})^{-1} \tilde{e^{a_1 \ldots a_m}} e^{b_1}\ldots e^{b_k}=\nonumber\\
&=&(e^{b_k})^{-1}\ldots (e^{b_1})^{-1} e^{a_m}\ldots e^{a_1} e^{b_1}\ldots e^{b_k}=\nonumber\\
&=&(-1)^{mk-j} e^{a_m}\ldots e^{a_1}=(-1)^{m-j} e^{a_m}\ldots e^{a_1}\nonumber
\end{eqnarray}
where $j$ is the number of elements $e^{a_1}, \ldots, e^{a_m}$ that have symmetric matrix representation. We can analogously prove remaining three formulas from (\ref{sogltransp}).

But formulas (\ref{sogltransp}) are not valid in real Clifford algebra of signatures $p-q\equiv4, 5, 6\!\!\mod 8$ because of the properties of operation of transposition for quaternionic matrices: $(AB)^T\neq B^T A^T$ for quaternionic matrices $A$ and $B$.

So, it makes sense to use the notion of additional signature of real Clifford algebra only in the cases $p-q\equiv3,7\!\!\mod 8$. We don't want to obtain all possible values of additional signature, but we want to obtain some possible values. If we consider matrix representation $\beta$ from the Section \ref{sectmatr},  then in the case $(p,q)=(0,1)$ we have $(k,l)=(1,0)$. Note, that we use only transformations of types 1, 3 and 4 from Section \ref{sectmatr} to obtain matrix representations of other Clifford algebras of signatures $p-q\equiv3,7\!\!\mod 8$. If we use transformation of type 1, then
$(p,q)\to(p+1, q+1)$ and $(k,l)\to(k+1, l+1)$. When we use transformation of type 3, we have $(p,q)\to(q+1, p-1)$ and $(k,l)\to(l+1, k-1)$ because $\beta^1$ is a symmetric matrix. When we use transformation of type 4, we have $(p,q)\to(p-4,q+4)$ and it is not difficult to obtain that $k$ and $l$ do not change the parity. So, for matrix representation $\beta$ we always have: if $p$ is even and $q$ is odd, then $k$ is odd and $l$ is even (note, that the same is true in complexified Clifford algebra, see table above). For example, for $(p,q)=(1,2)$ we have $(k,l)=(2,1)$, for $(p,q)=(3,0)$ we have $(k,l)=(2,1)$, for $(p,q)=(2,3)$ we have $(k,l)=(3,2)$ and so on. We will use this fact in the proof of Theorem 6.2 in the following section.

\section{Theorems}\label{sectmain}

Let us consider the following subsets of complexified Clifford algebra \cite{first}
\begin{eqnarray}
\G^{2i1}_{p,q} &=& \{U\in\cl^{(0)}_{p,q}\oplus i\cl^{(1)}_{p,q} : U^\ddagger U=e\},\nonumber\\
\G^{2i3}_{p,q} &=& \{U\in\cl^{(0)}_{p,q}\oplus i\cl^{(1)}_{p,q} : \hat{U}^{\ddagger} U=e\},\nonumber\\
\G^{23}_{p,q}&=&\{U\in\cl_{p,q}\, | \, \tilde{U}U=e\},\label{grou}\\
\G^{12}_{p,q}&=&\{U\in\cl_{p,q}\, | \, \tilde{\hat{U}}U=e\},\nonumber\\
\G^{2}_{p,q}&=&\{U\in\cl^{(0)}_{p,q}\, | \, \tilde{U}U=e\}.\nonumber
\end{eqnarray}
They can be considered as Lie groups. Their Lie algebras are
\begin{eqnarray}
\overline{\textbf{2}}\oplus i\overline{\textbf{1}},\quad \overline{\textbf{2}}\oplus i\overline{\textbf{3}},\quad \overline{\textbf{2}}\oplus\overline{\textbf{3}},\quad \overline{\textbf{2}}\oplus\overline{\textbf{1}},\quad \overline{\textbf{2}}\label{algebr}
\end{eqnarray}
respectively.

Note, that we have the following properties (see \cite{QuatAaca}, \cite{Quat2Aaca}, \cite{DAN})
\begin{eqnarray}
&&[\overline{\textbf{k}},\overline{\textbf{k}}]\subseteq\overline{\textbf{2}},\qquad k=0, 1, 2, 3 \nonumber;\\
&&[\overline{\textbf{k}},\overline{\textbf{2}}]\subseteq\overline{\textbf{k}}, \qquad k=0, 1, 2, 3 \label{1}; \\
&&[\overline{\textbf{0}},\overline{\textbf{1}}]\subseteq\overline{\textbf{3}}, \quad  [\overline{\textbf{0}},\overline{\textbf{3}}]\subseteq\overline{\textbf{1}}, \quad [\overline{\textbf{1}},\overline{\textbf{3}}]\subseteq\overline{\textbf{0}} \nonumber,
\end{eqnarray}
where $[U,V]=UV-VU$ is the commutator of arbitrary
Clifford algebra elements $U$ and $V$.

It is not difficult to calculate dimensions of these Lie algebras, because we know that $\dim \cl^k_{p,q}=C_n^k=\frac{n!}{k! (n-k)!}$. So, for example, $$\dim \overline{\textbf{2}}=\sum_{k\equiv 2\!\!\mod 4}C_n^k=2^{n-2}-2^{\frac{n-2}{2}}\cos{\frac{\pi n}{4}}.$$
Let us represent Lie groups, corresponding Lie algebras, and their dimensions in the following table.

\begin{center}
    \begin{tabular}{|c|c|c|}
\hline
Lie group & Lie algebra & dimension   \\ \hline \hline
$\G^{2i1}_{p,q}$ & $\overline{\textbf{2}}\oplus i\overline{\textbf{1}}$ & $2^{n-1}-2^{\frac{n-1}{2}}\cos{\frac{\pi (n+1)}{4}}$  \\ \hline
$\G^{2i3}_{p,q}$ & $\overline{\textbf{2}}\oplus i\overline{\textbf{3}}$ & $2^{n-1}-2^{\frac{n-1}{2}}\sin{\frac{\pi (n+1)}{4}}$   \\ \hline
$\G^{23}_{p,q}$ & $\overline{\textbf{2}}\oplus\overline{\textbf{3}}$ & $2^{n-1}-2^{\frac{n-1}{2}}\sin{\frac{\pi (n+1)}{4}}$  \\ \hline
$\G^{12}_{p,q}$ & $\overline{\textbf{2}}\oplus\overline{\textbf{1}}$ & $2^{n-1}-2^{\frac{n-1}{2}}\cos{\frac{\pi (n+1)}{4}}$  \\ \hline
$\G^{2}_{p,q}$ & $\overline{\textbf{2}}$ & $2^{n-2}-2^{\frac{n-2}{2}}\cos{\frac{\pi n}{4}}$  \\ \hline
\end{tabular}
\end{center}

\begin{thm}\label{theoremggg}\cite{first} We have the following Lie group isomorphisms
$$\G^{2i1}_{p,q}\simeq \G^{12}_{q,p},\qquad \G^{2i3}_{p,q}\simeq \G^{23}_{q,p},\qquad \G^{2}_{p,q}\simeq \G^{12}_{p,q-1}\simeq \G^{12}_{q,p-1},\qquad \G^{2}_{p,q}\simeq \G^{2}_{q,p}.$$
\end{thm}

\proof We must use transformation $e^a \to e^a e^n$ or $e^a \to ie^a$, $a=1, 2, \ldots, n$ in different cases. $\blacksquare$

According to Theorem \ref{theoremggg} it is sufficient to consider only groups $\G^{12}_{p,q}$ and $\G^{23}_{p,q}$.
In \cite{first} we proved isomorphisms between these groups and classical matrix Lie groups in the cases of signatures $p-q\equiv0, 1, 2\!\!\mod 8$. Now let us consider cases $p-q\equiv3, 7\!\!\mod 8$ and $p-q\equiv4, 5, 6\!\!\mod 8$.

\begin{thm} We have the following Lie group isomorphisms.
In the cases of signatures $p-q\equiv 3, 7\!\!\mod 8$ ($n$ is odd)
\begin{equation}
\G^{23}_{p,q}\simeq\left\lbrace
\begin{array}{ll}
\U(2^{\frac{n-1}{2}}), & \mbox{$(p,q)=(n,0)$;}\\
\U(2^{\frac{n-3}{2}},2^{\frac{n-3}{2}}), & \parbox{.5\linewidth}{$n\equiv 3, 7\!\!\mod 8$, $q\neq 0$;}\\
\Sp(2^{\frac{n-3}{2}},\C), & \parbox{.5\linewidth}{$n\equiv 5\!\!\mod 8$;}\\
\OO(2^{\frac{n-1}{2}}, \C), & \parbox{.5\linewidth}{$n\equiv 1\!\!\mod 8$.}
\end{array}
\right.\nonumber
\end{equation}
In the cases of signatures $p-q\equiv 3, 7\!\!\mod 8$ ($n$ is odd)
\begin{equation}
\G^{12}_{p,q}\simeq\left\lbrace
\begin{array}{ll}
\U(2^{\frac{n-1}{2}}), & \mbox{$(p,q)=(0,n)$;}\\
\U(2^{\frac{n-3}{2}},2^{\frac{n-3}{2}}), & \parbox{.5\linewidth}{$n\equiv 1, 5\!\!\mod 8$, $p\neq 0$;}\\
\Sp(2^{\frac{n-3}{2}},\C), & \parbox{.5\linewidth}{$n\equiv 3\!\!\mod 8$;}\\
\OO(2^{\frac{n-1}{2}}, \C), & \parbox{.5\linewidth}{$n\equiv 7\!\!\mod 8$.}
\end{array}
\right.\nonumber
\end{equation}
In the cases of signatures $p-q\equiv1, 5\!\!\mod 8$ ($n$ is odd)
\begin{equation}
\G^{2i1}_{p,q}\simeq\left\lbrace
\begin{array}{ll}
\U(2^{\frac{n-1}{2}}), & \mbox{$(p,q)=(n,0)$;}\\
\U(2^{\frac{n-3}{2}},2^{\frac{n-3}{2}}), & \parbox{.5\linewidth}{$n\equiv 1, 5\!\!\mod 8$, $q\neq 0$;}\\
\Sp(2^{\frac{n-3}{2}},\C), & \parbox{.5\linewidth}{$n\equiv 3\!\!\mod 8$;}\\
\OO(2^{\frac{n-1}{2}}, \C), & \parbox{.5\linewidth}{$n\equiv 7\!\!\mod 8$.}
\end{array}
\right.\nonumber
\end{equation}
In the cases of signatures $p-q\equiv1, 5\!\!\mod 8$ ($n$ is odd)
\begin{equation}
\G^{2i3}_{p,q}\simeq\left\lbrace
\begin{array}{ll}
\U(2^{\frac{n-1}{2}}), & \mbox{$(p,q)=(0,n)$;}\\
\U(2^{\frac{n-3}{2}},2^{\frac{n-3}{2}}), & \parbox{.5\linewidth}{$n\equiv 3, 7\!\!\mod 8$, $p\neq 0$;}\\
\Sp(2^{\frac{n-3}{2}},\C), & \parbox{.5\linewidth}{$n\equiv 5\!\!\mod 8$;}\\
\OO(2^{\frac{n-1}{2}}, \C), & \parbox{.5\linewidth}{$n\equiv 1\!\!\mod 8$.}
\end{array}
\right.\nonumber
\end{equation}
In the cases of signatures $p-q\equiv2, 6\!\!\mod 8$ ($n$ is even)
\begin{equation}
\G^{2}_{p,q}\simeq\left\lbrace
\begin{array}{ll}
\U(2^{\frac{n-2}{2}}), & \mbox{$(p,q)=(0,n), (0,n)$;}\\
\U(2^{\frac{n-4}{2}},2^{\frac{n-4}{2}}), & \parbox{.5\linewidth}{$n\equiv 2, 6\!\!\mod 8$, $p, q\neq 0$;}\\
\Sp(2^{\frac{n-4}{2}},\C), & \parbox{.5\linewidth}{$n\equiv 4\!\!\mod 8$;}\\
\OO(2^{\frac{n-2}{2}}, \C), & \parbox{.5\linewidth}{$n\equiv 0\!\!\mod 8$.}
\end{array}
\right.\nonumber
\end{equation}
\end{thm}
We use the standard notation of classical matrix groups \cite{Cornw}
\begin{eqnarray}
\U(n)&=&\{A\in\Mat(n,\C): A^\dagger A={\bf 1}\},\nonumber\\
\U(p,q)&=&\{A\in\Mat(p+q,\C): A^\dagger \eta A=\eta\},\nonumber\\
\Sp(n,\C)&=&\{A\in\Mat(2n,\C): A^T \Omega A=\Omega\},\nonumber\\
\OO(n,\C)&=&\{A\in\Mat(n,\C): A^T A={\bf 1}\},\nonumber
\end{eqnarray}
where $\Omega$ is the block matrix $\Omega=\left( \begin{array}{ll}
 0 & -{\bf1} \\
 {\bf1} & 0 \end{array}\right)$.

\proof We consider Lie group $\G^{23}_{p,q}$ in Clifford algebra $\cl_{p,q}\simeq \Mat(\frac{n-1}{2},\C)$, $p-q\equiv3, 7\!\!\mod 8$. We use complex matrix representation $\beta: U\to \beta(U)$, $U\in\cl_{p,q}$ in the cases $p-q\equiv3, 7\!\!\mod 8$ from Section \ref{sectmatr} such that
$\beta(U^\dagger)=(\beta(U))^\dagger$.

In the case of signature $(n,0)$ we have $\tilde{U}=U^\dagger$ (see (\ref{1rt})). From $\tilde{U} U=e$ we obtain $U^\dagger U={\bf 1}$ and isomorphism with group $\U(2^{\frac{n-1}{2}})$.

Consider the case $q\neq 0$. If $p$ is odd and $q$ is even, then $\tilde{U}=e_{1\ldots p}U^\dagger e^{1\ldots p}$. In the case $p\equiv1\!\!\mod 4$ (in this case we have $q\equiv2\!\!\mod 4$ and $n\equiv3\!\!\mod 4$) let us consider matrix $M=\beta^{1\ldots p}$. We have $M^2=(-1)^{\frac{p(p-1)}{2}}{\bf 1}={\bf 1}$, $M^\dagger=M^{-1}$, $\tr M=0$. Thus, the spectrum of matrix $M$ consists of the same number of $1$ and $-1$. So there exists such unitary matrix $T^\dagger=T^{-1}$ that $J=T^{-1}MT$, where $J=\diag(1, \ldots, 1, -1, \ldots ,-1)$ is the diagonal matrix with the same number of $1$ and $-1$ on the diagonal. Now we consider transformation $T^{-1}\beta^a T=\gamma^a$ and obtain another matrix representation $\gamma$ of Clifford algebra with $\gamma^{1\ldots p}=J$. From $\tilde{U} U=e$ we obtain $U^\dagger J U = J$ and isomorphism with the group $\U(2^{\frac{n-3}{2}},2^{\frac{n-3}{2}})$. The case $p\equiv3\!\!\mod 4$ (in this case we have $q\equiv0\!\!\mod 4$, $n\equiv3\!\!\mod 4$) is similar (we must consider element $M=i \beta^{1\ldots p}$).

Now let $p$ be even and $q$ be odd. Then $n\equiv1\!\!\mod 4$. We know that $k$ is odd and $l$ is even in these cases (see previous section). So we can use formula $U^\T=e_{b_1 \ldots b_k}\tilde{U} e^{b_1 \ldots b_k}$. Let us consider matrix $M=\beta^{b_1 \ldots b_k}$. We have $M^T=M^{-1}$, $\tr M=0$, $M^2={\bf 1}$ (or $M^2=-{\bf 1}$). Thus there exists such orthogonal matrix $T^T=T^{-1}$ that $J=T^{-1}MT$ (or $\Omega=T^{-1}MT$, because for $\Omega$ we also have $\Omega^T=\Omega$, $\Omega^2=-{\bf 1}$, $\tr\Omega=0$).
We consider transformation $T^{-1}\beta^a T=\gamma^a$ and obtain another matrix representation $\gamma$ of Clifford algebra with $\gamma^{b_1\ldots b_k}=J$ (or $\gamma^{b_1\ldots b_k}=\Omega$). So we obtain condition $U^T J U=J$ (or $U^T \Omega U=\Omega$). Thus we obtain in the case $n\equiv 1\!\!\mod 8$ isomorphism with group $\OO(2^{\frac{n-3}{2}},2^{\frac{n-3}{2}}, \C)\simeq\OO(2^{\frac{n-1}{2}},\C)$ and in the case $n\equiv 5\!\!\mod 8$ isomorphism with group $\Sp(2^{\frac{n-3}{2}},\C)$. We take into account that $\dim \G^{23}_{p,q}=2^{n-1}-2^{\frac{n-1}{2}}\sin{\frac{\pi (n+1)}{4}}$, $\dim \OO(2^{\frac{n-1}{2}},\C)=2^{n-1}-2^{\frac{n-1}{2}}$, $\dim \Sp(2^{\frac{n-3}{2}},\C)=2^{n-1}+2^{\frac{n-1}{2}}$.

We can obtain isomorphisms for the group $\G^{12}_{p,q}$ analogously. The statement for the groups $\G^{2i1}_{p,q}$, $\G^{2i3}_{p,q}$ and $\G^{2}_{p,q}$ follows from Theorem \ref{theoremggg}. $\blacksquare$

\begin{thm} We have the following Lie group isomorphisms.
In the cases of signatures $p-q\equiv4, 5, 6\!\!\mod 8$
\begin{equation}
\G^{23}_{p,q}\simeq\left\lbrace
\begin{array}{ll}
\Sp(2^{\frac{n-2}{2}}), & \mbox{$(p,q)=(n,0)$, $n$ is even;}\\
\Sp(2^{\frac{n-4}{2}}, 2^{\frac{n-4}{2}}), & \parbox{.5\linewidth}{$n\equiv 4, 6\!\!\mod 8$, $q\neq 0$;}\\
\OO(2^{\frac{n-2}{2}},\H), & \parbox{.5\linewidth}{$n\equiv 0, 2\!\!\mod 8$;}\\
\Sp(2^{\frac{n-3}{2}})\times \Sp(2^{\frac{n-3}{2}}), & \mbox{$(p,q)=(n,0)$, $n$ is odd;}\\
\Sp(2^{\frac{n-5}{2}}, 2^{\frac{n-5}{2}})\times \Sp(2^{\frac{n-5}{2}}, 2^{\frac{n-5}{2}}), & \parbox{.5\linewidth}{$n\equiv 5\!\!\mod 8$, $q\neq 0$;}\\
\OO(2^{\frac{n-3}{2}},\H)\times \OO(2^{\frac{n-3}{2}},\H), & \parbox{.5\linewidth}{$n\equiv 1\!\!\mod 8$;}\\
\GL(2^{\frac{n-3}{2}},\H), & \parbox{.5\linewidth}{$n\equiv 3, 7\!\!\mod 8$.}
\end{array}
\right.\nonumber
\end{equation}
In the cases of signatures $p-q\equiv4, 5, 6\!\!\mod 8$
\begin{equation}
\G^{12}_{p,q}\simeq\left\lbrace
\begin{array}{ll}
\Sp(2^{\frac{n-2}{2}}), & \mbox{$(p,q)=(0,n)$, $n$ is even;}\\
\Sp(2^{\frac{n-4}{2}}, 2^{\frac{n-4}{2}}), & \parbox{.5\linewidth}{$n\equiv 2, 4\!\!\mod 8$, $p\neq 0$;}\\
\OO(2^{\frac{n-2}{2}},\H), & \parbox{.5\linewidth}{$n\equiv 0, 6\!\!\mod 8$;}\\
\Sp(2^{\frac{n-3}{2}})\times \Sp(2^{\frac{n-3}{2}}), & \mbox{$(p,q)=(0,n)$, $n$ is odd;}\\
\Sp(2^{\frac{n-5}{2}}, 2^{\frac{n-5}{2}})\times \Sp(2^{\frac{n-5}{2}}, 2^{\frac{n-5}{2}}), & \parbox{.5\linewidth}{$n\equiv 3\!\!\mod 8$, $p\neq 0$;}\\
\OO(2^{\frac{n-3}{2}},\H)\times \OO(2^{\frac{n-3}{2}},\H), & \parbox{.5\linewidth}{$n\equiv 7\!\!\mod 8$;}\\
\GL(2^{\frac{n-3}{2}},\H), & \parbox{.5\linewidth}{$n\equiv 1, 5\!\!\mod 8$.}
\end{array}
\right.\nonumber
\end{equation}
In the cases of signatures $p-q\equiv2, 3, 4\!\!\mod 8$
\begin{equation}
\G^{2i1}_{p,q}\simeq\left\lbrace
\begin{array}{ll}
\Sp(2^{\frac{n-2}{2}}), & \mbox{$(p,q)=(n,0)$, $n$ is even;}\\
\Sp(2^{\frac{n-4}{2}}, 2^{\frac{n-4}{2}}), & \parbox{.5\linewidth}{$n\equiv 2, 4\!\!\mod 8$, $q\neq 0$;}\\
\OO(2^{\frac{n-2}{2}},\H), & \parbox{.5\linewidth}{$n\equiv 0, 6\!\!\mod 8$;}\\
\Sp(2^{\frac{n-3}{2}})\times \Sp(2^{\frac{n-3}{2}}), & \mbox{$(p,q)=(n,0)$, $n$ is odd;}\\
\Sp(2^{\frac{n-5}{2}}, 2^{\frac{n-5}{2}})\times \Sp(2^{\frac{n-5}{2}}, 2^{\frac{n-5}{2}}), & \parbox{.5\linewidth}{$n\equiv 3\!\!\mod 8$, $q\neq 0$;}\\
\OO(2^{\frac{n-3}{2}},\H)\times \OO(2^{\frac{n-3}{2}},\H), & \parbox{.5\linewidth}{$n\equiv 7\!\!\mod 8$;}\\
\GL(2^{\frac{n-3}{2}},\H), & \parbox{.5\linewidth}{$n\equiv 1, 5\!\!\mod 8$.}
\end{array}
\right.\nonumber
\end{equation}
In the cases of signatures $p-q\equiv2, 3, 4\!\!\mod 8$
\begin{equation}
\G^{2i3}_{p,q}\simeq\left\lbrace
\begin{array}{ll}
\Sp(2^{\frac{n-2}{2}}), & \mbox{$(p,q)=(0,n)$, $n$ is even;}\\
\Sp(2^{\frac{n-4}{2}}, 2^{\frac{n-4}{2}}), & \parbox{.5\linewidth}{$n\equiv 4, 6\!\!\mod 8$, $p\neq 0$;}\\
\OO(2^{\frac{n-2}{2}},\H), & \parbox{.5\linewidth}{$n\equiv 0, 2\!\!\mod 8$;}\\
\Sp(2^{\frac{n-3}{2}})\times \Sp(2^{\frac{n-3}{2}}), & \mbox{$(p,q)=(0,n)$, $n$ is odd;}\\
\Sp(2^{\frac{n-5}{2}}, 2^{\frac{n-5}{2}})\times \Sp(2^{\frac{n-5}{2}}, 2^{\frac{n-5}{2}}), & \parbox{.5\linewidth}{$n\equiv 5\!\!\mod 8$, $p\neq 0$;}\\
\OO(2^{\frac{n-3}{2}},\H)\times \OO(2^{\frac{n-3}{2}},\H), & \parbox{.5\linewidth}{$n\equiv 1\!\!\mod 8$;}\\
\GL(2^{\frac{n-3}{2}},\H), & \parbox{.5\linewidth}{$n\equiv 3, 7\!\!\mod 8$.}
\end{array}
\right.\nonumber
\end{equation}
In the cases of signatures $p-q\equiv3, 4, 5\!\!\mod 8$
\begin{equation}
\G^{2}_{p,q}\simeq\left\lbrace
\begin{array}{ll}
\Sp(2^{\frac{n-3}{2}}), & \mbox{$(p,q)=(n,0), (0,n)$, $n$ is odd;}\\
\Sp(2^{\frac{n-5}{2}}, 2^{\frac{n-5}{2}}), & \parbox{.5\linewidth}{$n\equiv 3, 5\!\!\mod 8$, $p, q\neq 0$;}\\
\OO(2^{\frac{n-3}{2}},\H), & \parbox{.5\linewidth}{$n\equiv 1, 7\!\!\mod 8$;}\\
\Sp(2^{\frac{n-4}{2}})\times \Sp(2^{\frac{n-4}{2}}), & \mbox{$(p,q)=(n,0), (0,n)$, $n$ is even;}\\
\Sp(2^{\frac{n-6}{2}}, 2^{\frac{n-6}{2}})\times \Sp(2^{\frac{n-6}{2}}, 2^{\frac{n-6}{2}}), & \parbox{.5\linewidth}{$n\equiv 4\!\!\mod 8$, $p, q\neq 0$;}\\
\OO(2^{\frac{n-4}{2}},\H)\times \OO(2^{\frac{n-4}{2}},\H), & \parbox{.5\linewidth}{$n\equiv 0\!\!\mod 8$;}\\
\GL(2^{\frac{n-4}{2}},\H), & \parbox{.5\linewidth}{$n\equiv 2, 6\!\!\mod 8$.}
\end{array}
\right.\nonumber
\end{equation}
\end{thm}
We use the standard notation of classical matrix groups \cite{Cornw}
\begin{eqnarray}
\GL(n,\H)&=&\{A\in\Mat(n,\H): \exists A^{-1} \},\nonumber\\
\Sp(n)&=&\{A\in\GL(n,\H): A^*A={\bf 1}\},\nonumber\\
\Sp(p,q)&=&\{A\in\GL(p+q,\H): A^* \eta A=\eta\},\nonumber\\
\OO(n,\H)&=&\OO^{*}(2n)=\{A\in\GL(n,\H): A^* i {\bf 1} A=i {\bf 1}\},\nonumber
\end{eqnarray}
where $*$ is the operation of conjugate transpose of matrix of quaternions.

\proof Let us consider Lie group $\G^{23}_{p,q}$. In the cases $p-q\equiv4, 6\!\!\mod 8$ we have $\cl_{p,q}\simeq \Mat(2^{\frac{n-2}{2}},\mathbb H)$. In the cases $p-q\equiv5\!\!\mod 8$ we have $\cl_{p,q}\simeq \Mat(2^{\frac{n-3}{2}},\mathbb H)\oplus \Mat(2^{\frac{n-3}{2}},\mathbb H)$ and we use block-diagonal representation $\beta$ of Clifford algebra (see Section \ref{sectmatr}).

In the case of signature $(n,0)$ we have $U^\dagger=\tilde{U}$ and obtain $U^\dagger U=e$. We have $(\beta(U))^*=\beta(U^\dagger)$, where $*$ is the operation of conjugate transpose of matrices of quaternions. So we obtain isomorphism with the group $\Sp(2^{\frac{n-2}{2}})$ in the case of even $n$ and isomorphism with the group $\Sp(2^{\frac{n-3}{2}})\times \Sp(2^{\frac{n-3}{2}})$ in the case of odd $n$.

Now let us consider the case $q\neq 0$. Let $n$ be even. If $p$ and $q$ are even, then we have $U^\dagger=e_{p+1 \ldots n}\tilde{U} e^{p+1 \ldots n}$ and obtain $U^\dagger e^{p+1 \ldots n} U= e^{p+1 \ldots n}$. If $p$ and $q$ are odd, then we have $U^\dagger=e_{1 \ldots p}\tilde{U}e^{1 \ldots p}$
and obtain $U^\dagger e^{1 \ldots p} U= e^{1 \ldots p}$. Let us consider matrix $M=\beta^{p+1 \ldots n}$ (or $\beta^{1\ldots p}$ in the corresponding case). We have $M^*=M^{-1}$ and $M^2={\bf \pm 1}$, so $M^*=\pm M$. It is known (see \cite{quaternion}) that any square quaternionic matrix of size $m$ has the Jordan canonical form with (standard) $m$ right eigenvalues (which are complex numbers with nonnegative imaginary parts) on the diagonal. If matrix is normal ($A^*A=A A^*$), then there exists a unitary matrix $T^*=T^{-1}$ such that $T^* A T$ is a diagonal matrix with standard right eigenvalues on the diagonal. If this  matrix is Hermitian, then these right eigenvalues are real. We obtain that there exists such element $T^*=T^{-1}$ that $\eta=T^{-1}MT$ or $i{\bf 1}=T^{-1}MT$. We use transformation $T^{-1}\beta^a T=\gamma^a$ and obtain another matrix representation $\gamma$ of Clifford algebra with $\gamma^{p+1\ldots n}=\eta$ or $i{\bf 1}$ (or analogously for element $\gamma^{1\ldots p}$). We obtain $ U^* \eta U=\eta$ or $U^* i {\bf 1} U=i {\bf 1}$ and isomorphisms with group $\Sp(2^{\frac{n-4}{2}}, 2^{\frac{n-4}{2}})$ or $\OO(2^{\frac{n-2}{2}},\H)$. We take into account that $\dim \G^{23}_{p,q}=2^{n-1}-2^{\frac{n-1}{2}}\sin{\frac{\pi (n+1)}{4}}$, $\dim \Sp(2^{\frac{n-4}{2}},2^{\frac{n-4}{2}})=2^{n-1}+2^{\frac{n-2}{2}}$ and $\dim \OO(2^{\frac{n-2}{2}},\H)=2^{n-1}-2^{\frac{n-2}{2}}$.

If $n\equiv 1, 5\!\!\mod 8$, then $p$ is odd and $q$ is even. So we can use formula $U^\dagger=e_{p+1 \ldots n}\tilde{U}e^{p+1 \ldots n}$ and obtain $U^\dagger e^{p+1 \ldots n} U= e^{p+1 \ldots n}$. Our considerations are like considerations in the case of even $n$ but our matrix representation is block-diagonal. We obtain isomorphisms with groups $\OO(2^{\frac{n-3}{2}},\H)\times \OO(2^{\frac{n-3}{2}},\H)$ or $\Sp(2^{\frac{n-5}{2}}, 2^{\frac{n-5}{2}})\times \Sp(2^{\frac{n-5}{2}}, 2^{\frac{n-5}{2}})$ in different cases.

If $n\equiv 3, 7\!\!\mod 8$, then $p$ is even and $q$ is odd. So, we have $U^*=e_{1\ldots p}\tilde{\hat{U}}e^{1 \ldots p}$ and $\tilde{U}=(e^{1\ldots p})^{-1}\hat{U}^{*}e^{1\ldots p}$. We have block-diagonal matrix representation $\beta$ (see Section \ref{sectmatr}). The even part of an arbitrary Clifford algebra element has the form $\diag(A,A)$ and the odd part of an element has the form $\diag(B,-B)$. Then we obtain from $\tilde{U} U=e$:
$$(\diag(A-B, A+B))^* \diag(G,G)\diag(A+B,A-B)=\diag(G,G),$$
where $\diag(G,G)$ is matrix representation of element $e^{1\ldots p}$. So, we obtain $(A-B)^T G(A+B)=G$ and isomorphism with group $\GL(2^{\frac{n-3}{2}},\H)$ because for any invertible matrix $A$ there exists such matrix $B$ that $(A-B)^T G(A+B)=G$. Note, that in particular case $p=0$ we have $U^*=\tilde{\hat{U}}$ and we obtain $\hat{U}^{*}U=e$, $(\diag(A-B, A+B))^*\diag(A+B,A-B)={\bf 1}$, $(A-B)^T(A+B)={\bf 1}$,  and isomorphism with linear group again.

We can obtain isomorphisms for the group $\G^{12}_{p,q}$ analogously. The statement for the groups $\G^{2i1}_{p,q}$, $\G^{2i3}_{p,q}$ and $\G^{2}_{p,q}$ follows from Theorem \ref{theoremggg}. $\blacksquare$

\section{Relation between group $\G^{2}_{p,q}$ and spin group}
\label{SectionSpin}

Note that spin group \cite{Lounesto}, \cite{spin}
$$\Spin_{+}(p,q)=\{U\in\cl^{(0)}_{p,q} | \forall x\in\cl^1_{p,q}, UxU^{-1}\in\cl^1_{p,q}, \tilde{U}U=e\}$$
is a subgroup of all five considered Lie groups (\ref{grou}). Moreover, group $\Spin_{+}(p,q)$ coincides with group
$$\G^{2}_{p,q}=\{U\in\cl^{(0)}_{p,q}\, | \, \tilde{U}U=e\}$$
in the cases of dimensions $n\leq 5$. Lie algebra $\cl^2_{p,q}$ of Lie group $\Spin_{+}(p,q)$ is a subalgebra of algebras (\ref{algebr}). Moreover, Lie algebra $\cl^2_{p,q}$ coincides with Lie algebra $\overline{\textbf{2}}$ in the cases of dimensions $n\leq 5$, because the notion of grade 2 and the notion of quaternion type 2 coincide in these cases.

Let us represent all isomorphisms of the group $\G^{2}_{p,q}$ in the following endless table. We use notation $^2\Sp(1)=\Sp(1)\times\Sp(1)$ and similar notation.

\begin{center}
{\footnotesize
\begin{center}
\begin{tabular}{|c||c|c|c|c|c|c|c|c|}
\hline
$p \diagdown q$ & 0       & 1          & 2            &  3          & 4          & 5             & 6            & 7           \\ \hline \hline
0   & $\OO(1)$ & $\OO(1)$    & $\U(1)$      &  $\Sp(1)$  & $^2\Sp(1)$  &   $\Sp(2)$  & $\U(4)$      & $\OO(8)$      \\ \hline
1   & $\OO(1)$ & $\GL(1,\R)$& $\Sp(1,\R)$  & $\Sp(1,\C)$ & $\Sp(1,1)$ & $\GL(2,\H)$& $\OO(4,\H)$ & $\OO(8,\C)$ \\ \hline
2   & $\U(1)$ & $\Sp(1,\R)$& $^2\Sp(1,\R)$& $\Sp(2,\R)$ & $\U(2,2)$  & $\OO(4,\H)$   & $^2\OO(4,\H)$  & $\OO(8,\H)$     \\ \hline
3   & $\Sp(1)$& $\Sp(1,\C)$& $\Sp(2,\R)$  & $\GL(4,\R)$ & $\OO(4,4)$  & $\OO(8,\C)$   &$\OO(8,\H)$ & $\GL(8,\H)$       \\ \hline
4   &$^2\Sp(1)$ & $\Sp(1,1)$ & $\U(2,2)$    & $\OO(4,4)$   &$^2\OO(4,4)$ & $\OO(8,8)$     & $\U(8,8)$    & $\Sp(8,8)$  \\ \hline
5   & $\Sp(2)$&$\GL(2,\H)$ & $\OO(4,\H)$  &$\OO(8,\C)$  & $\OO(8,8)$  & $\GL(16,\R)$  & $\Sp(16,\R)$ &$\Sp(16,\C)$     \\ \hline
6   & $\U(4)$ &$\OO(4,\H)$ & $^2\OO(4,\H)$&  $\OO(8,\H)$ & $\U(8,8)$  &$\Sp(16,\R)$   &$^2\Sp(16,\R)$&$\Sp(32,\R)$\\ \hline
7   &$\OO(8)$  & $\OO(8,\C)$&  $\OO(8,\H)$   & $\GL(8,\H)$ & $\Sp(8,8)$ &$\Sp(16,\C)$    &$\Sp(32,\R)$  &$\GL(64,\R)$    \\ \hline
\end{tabular}
\end{center}
}
\end{center}

Note that isomorphisms of the group $\Spin_{+}(p,q)$ are well-known (see, for example, \cite{Lounesto}, table on the page 224):

\begin{center}
\begin{tabular}{|c||c|c|c|c|c|c|c|}
\hline
$p \diagdown q$ & 0 & 1 & 2 & 3 & 4 & 5 & 6   \\ \hline \hline
0 & $\OO(1)$ & $\OO(1)$ & $\U(1)$ & $\SU(2)$  & $ ^2\SU(2)$ & $\Sp(2)$ & $\SU(4)$ \\ \hline
1 & $\OO(1)$& $\GL(1,\R)$& $\Sp(1,\R)$ & $\Sp(1,\C)$ & $\Sp(1,1)$ & $\SL(2,\H)$ &  \\ \hline
2 & $\U(1)$ & $\Sp(1,\R)$ & $^2\Sp(1,\R)$ & $\Sp(2,\R)$ & $\SU(2,2)$ &  &   \\ \hline
3 & $\SU(2)$ & $\Sp(1,\C)$ & $\Sp(2,\R)$ & $\SL(4,\R)$  &  &  &   \\ \hline
4 & $ ^2\SU(2)$& $\Sp(1,1)$ & $\SU(2,2)$ &  &  &  &   \\ \hline
5 & $\Sp(2)$ & $\SL(2,\H)$ &  &  &  &  &   \\ \hline
6 & $\SU(4)$ &  &  &  &  &  &   \\ \hline
\end{tabular}
\end{center}

So, in the cases $n\leq 5$ the tables coincide (note, that $\SU(2)\simeq \Sp(1)$). In the cases $n=6$ condition $U^{-1}\cl^1_{p,q}U\in\cl^1_{p,q}$ in the definition of the group $\Spin_{+}(p,q)$ transforms into condition $\det \gamma(U)=1$ for matrix representation $\gamma$ and we obtain special groups.

Note that group $\Spin_{+}(p,q)$ in the cases $n\geq 7$ is not directly related to classical matrix groups (see \cite{Lounesto}, p.224).
But we present classical matrix groups that contain this group in the cases of arbitrary dimension $n\geq 7$ and signature $(p,q)$.

\section{Conclusion}\label{sectconcl}

Using results of the present paper and \cite{first}, we can represent isomorphisms between considered five Lie groups and classical matrix groups in the following tables. There is $n\!\!\mod 8$ in the lines and there is $p-q\!\!\mod 8$ in the columns.

$$\G^{12}_{p,q}=\{U\in\cl_{p,q}\, | \, \tilde{\hat{U}}U=e\}:$$
\begin{center}
\begin{tabular}{|c||c|c|}
\hline
$n \diagdown p-q$ & $0, 2$              & $4, 6$    \\ \hline \hline
$0, 6$   & $\begin{array}{l} \OO(2^{\frac{n-2}{2}},2^{\frac{n-2}{2}}), p\neq 0;\\ \OO(2^{\frac{n}{2}}), p=0. \end{array}$

         & $\OO(2^{\frac{n-2}{2}},\H)$    \\ \hline
$2, 4$   & $\Sp(2^{\frac{n-2}{2}},\R)$

         & $\begin{array}{l} \Sp(2^{\frac{n-4}{2}},2^{\frac{n-4}{2}}), p\neq 0;\\ \Sp(2^\frac{n-2}{2}), p=0. \end{array}$       \\ \hline
\end{tabular}
\end{center}

\begin{center}
\begin{tabular}{|c||c|c|c|}
\hline
$n \diagdown p-q$ & $1$       & $3, 7$         & $5$    \\ \hline \hline
$7$      & $\begin{array}{l} ^2\OO(2^{\frac{n-3}{2}},2^{\frac{n-3}{2}}), p\neq 0;\\ ^2\OO(2^{\frac{n-1}{2}}), p=0. \end{array}$
         & $\OO(2^{\frac{n-1}{2}},\C)$
         & $^2\OO(2^{\frac{n-3}{2}},\H)$    \\ \hline
$3$      & $^2\Sp(2^{\frac{n-3}{2}},\R)$
         & $\Sp(2^{\frac{n-3}{2}},\C)$
         & $\begin{array}{l} ^2\Sp(2^{\frac{n-5}{2}},2^{\frac{n-5}{2}}), p\neq 0;\\ ^2\Sp(2^{\frac{n-3}{2}}), p=0. \end{array}$      \\ \hline
$1, 5$   & $\GL(2^{\frac{n-1}{2}},\R)$
         & $\begin{array}{l} \U(2^{\frac{n-3}{2}},2^{\frac{n-3}{2}}), p\neq 0;\\ \U(\frac{n-1}{2}), p=0. \end{array}$
         & $\GL(2^{\frac{n-3}{2}},\H)$    \\ \hline
\end{tabular}
\end{center}

$$\G^{23}_{p,q}=\{U\in\cl_{p,q}\, | \, \tilde{U}U=e\}:$$
\begin{center}
\begin{tabular}{|c||c|c|}
\hline
$n \diagdown p-q$ & $0, 2$              & $4, 6$    \\ \hline \hline
$0, 2$   & $\begin{array}{l} \OO(2^{\frac{n-2}{2}},2^{\frac{n-2}{2}}), q\neq 0;\\ \OO(2^{\frac{n}{2}}), q=0. \end{array}$

         & $\OO(2^{\frac{n-2}{2}},\H)$    \\ \hline
$4, 6$   & $\Sp(2^{\frac{n-2}{2}},\R)$

         & $\begin{array}{l} \Sp(2^{\frac{n-4}{2}},2^{\frac{n-4}{2}}), q\neq 0;\\ \Sp(2^\frac{n-2}{2}), q=0. \end{array}$       \\ \hline
\end{tabular}
\end{center}

\begin{center}
\begin{tabular}{|c||c|c|c|}
\hline
$n \diagdown p-q$ & $1$       & $3, 7$         & $5$    \\ \hline \hline
$1$      & $\begin{array}{l} ^2\OO(2^{\frac{n-3}{2}},2^{\frac{n-3}{2}}), q\neq 0;\\ ^2\OO(2^{\frac{n-1}{2}}), q=0. \end{array}$
         & $\OO(2^{\frac{n-1}{2}},\C)$
         & $^2\OO(2^{\frac{n-3}{2}},\H)$    \\ \hline
$5$      & $^2\Sp(2^{\frac{n-3}{2}},\R)$
         & $\Sp(2^{\frac{n-3}{2}},\C)$
         & $\begin{array}{l} ^2\Sp(2^{\frac{n-5}{2}},2^{\frac{n-5}{2}}), q\neq 0;\\ ^2\Sp(2^{\frac{n-3}{2}}), q=0. \end{array}$      \\ \hline
$3, 7$   & $\GL(2^{\frac{n-1}{2}},\R)$
         & $\begin{array}{l} \U(2^{\frac{n-3}{2}},2^{\frac{n-3}{2}}), q\neq 0;\\ \U(\frac{n-1}{2}), q=0. \end{array}$
         & $\GL(2^{\frac{n-3}{2}},\H)$    \\ \hline
\end{tabular}
\end{center}

$$\G^{2i1}_{p,q}=\{U\in\cl^{(0)}_{p,q}\oplus i\cl^{(1)}_{p,q} : U^\ddagger U=e\}:$$
\begin{center}
\begin{tabular}{|c||c|c|}
\hline
$n \diagdown p-q$ & $0, 6$              & $2, 4$    \\ \hline \hline
$0, 6$   & $\begin{array}{l} \OO(2^{\frac{n-2}{2}},2^{\frac{n-2}{2}}), q\neq 0;\\ \OO(2^{\frac{n}{2}}), q=0. \end{array}$

         & $\OO(2^{\frac{n-2}{2}},\H)$    \\ \hline
$2, 4$   & $\Sp(2^{\frac{n-2}{2}},\R)$

         & $\begin{array}{l} \Sp(2^{\frac{n-4}{2}},2^{\frac{n-4}{2}}), q\neq 0;\\ \Sp(2^\frac{n-2}{2}), q=0. \end{array}$       \\ \hline
\end{tabular}
\end{center}

\begin{center}
\begin{tabular}{|c||c|c|c|}
\hline
$n \diagdown p-q$ & $7$       & $1, 5$         & $3$    \\ \hline \hline
$7$      & $\begin{array}{l} ^2\OO(2^{\frac{n-3}{2}},2^{\frac{n-3}{2}}), q\neq 0;\\ ^2\OO(2^{\frac{n-1}{2}}), q=0. \end{array}$
         & $\OO(2^{\frac{n-1}{2}},\C)$
         & $^2\OO(2^{\frac{n-3}{2}},\H)$    \\ \hline
$3$      & $^2\Sp(2^{\frac{n-3}{2}},\R)$
         & $\Sp(2^{\frac{n-3}{2}},\C)$
         & $\begin{array}{l} ^2\Sp(2^{\frac{n-5}{2}},2^{\frac{n-5}{2}}), q\neq 0;\\ ^2\Sp(2^{\frac{n-3}{2}}), q=0. \end{array}$      \\ \hline
$1, 5$   & $\GL(2^{\frac{n-1}{2}},\R)$
         & $\begin{array}{l} \U(2^{\frac{n-3}{2}},2^{\frac{n-3}{2}}), q\neq 0;\\ \U(\frac{n-1}{2}), q=0. \end{array}$
         & $\GL(2^{\frac{n-3}{2}},\H)$    \\ \hline
\end{tabular}
\end{center}

$$\G^{2i3}_{p,q}=\{U\in\cl^{(0)}_{p,q}\oplus i\cl^{(1)}_{p,q} : \hat{U}^{\ddagger} U=e\}:$$
\begin{center}
\begin{tabular}{|c||c|c|}
\hline
$n \diagdown p-q$ & $0, 6$              & $2, 4$    \\ \hline \hline
$0, 2$   & $\begin{array}{l} \OO(2^{\frac{n-2}{2}},2^{\frac{n-2}{2}}), p\neq 0;\\ \OO(2^{\frac{n}{2}}), p=0. \end{array}$

         & $\OO(2^{\frac{n-2}{2}},\H)$    \\ \hline
$4, 6$   & $\Sp(2^{\frac{n-2}{2}},\R)$

         & $\begin{array}{l} \Sp(2^{\frac{n-4}{2}},2^{\frac{n-4}{2}}), p\neq 0;\\ \Sp(2^\frac{n-2}{2}), p=0. \end{array}$       \\ \hline
\end{tabular}
\end{center}

\begin{center}
\begin{tabular}{|c||c|c|c|}
\hline
$n \diagdown p-q$ & $7$       & $1, 5$         & $3$    \\ \hline \hline
$1$      & $\begin{array}{l} ^2\OO(2^{\frac{n-3}{2}},2^{\frac{n-3}{2}}), p\neq 0;\\ ^2\OO(2^{\frac{n-1}{2}}), p=0. \end{array}$
         & $\OO(2^{\frac{n-1}{2}},\C)$
         & $^2\OO(2^{\frac{n-3}{2}},\H)$    \\ \hline
$5$      & $^2\Sp(2^{\frac{n-3}{2}},\R)$
         & $\Sp(2^{\frac{n-3}{2}},\C)$
         & $\begin{array}{l} ^2\Sp(2^{\frac{n-5}{2}},2^{\frac{n-5}{2}}), p\neq 0;\\ ^2\Sp(2^{\frac{n-3}{2}}), p=0. \end{array}$      \\ \hline
$3, 7$   & $\GL(2^{\frac{n-1}{2}},\R)$
         & $\begin{array}{l} \U(2^{\frac{n-3}{2}},2^{\frac{n-3}{2}}), p\neq 0;\\ \U(\frac{n-1}{2}),p=0. \end{array}$
         & $\GL(2^{\frac{n-3}{2}},\H)$    \\ \hline
\end{tabular}
\end{center}

$$\G^{2}_{p,q}=\{U\in\cl^{(0)}_{p,q}\, | \, \tilde{U}U=e\}:$$
\begin{center}
\begin{tabular}{|c||c|c|}
\hline
$n \diagdown p-q$ & $1, 7$              & $3, 5$    \\ \hline \hline
$1, 7$   & $\begin{array}{l} \OO(2^{\frac{n-3}{2}},2^{\frac{n-3}{2}}), p, q\neq 0;\\ \OO(2^{\frac{n-1}{2}}), (n,0), (0,n). \end{array}$

         & $\OO(2^{\frac{n-3}{2}},\H)$    \\ \hline
$3, 5$   & $\Sp(2^{\frac{n-3}{2}},\R)$

         & $\begin{array}{l} \Sp(2^{\frac{n-5}{2}},2^{\frac{n-5}{2}}), p, q\neq 0;\\ \Sp(2^\frac{n-3}{2}), (n,0), (0,n). \end{array}$       \\ \hline
\end{tabular}
\end{center}
%
\begin{center}
\begin{tabular}{|c||c|c|c|}
\hline
$n \diagdown p-q$ & $0$       & $2, 6$         & $4$    \\ \hline \hline
$0$      & $\begin{array}{l} ^2\OO(2^{\frac{n-4}{2}},2^{\frac{n-4}{2}}), p, q\neq 0;\\ ^2\OO(2^{\frac{n-2}{2}}), (n,0), (0,n). \end{array}$
         & $\OO(2^{\frac{n-2}{2}},\C)$
         & $^2\OO(2^{\frac{n-4}{2}},\H)$    \\ \hline
$4$      & $^2\Sp(2^{\frac{n-4}{2}},\R)$
         & $\Sp(2^{\frac{n-4}{2}},\C)$
         & $\begin{array}{l} ^2\Sp(2^{\frac{n-6}{2}},2^{\frac{n-6}{2}}), p, q\neq 0;\\ ^2\Sp(2^{\frac{n-4}{2}}), (n,0), (0,n). \end{array}$      \\ \hline
$2, 6$   & $\GL(2^{\frac{n-2}{2}},\R)$
         & $\begin{array}{l} \U(2^{\frac{n-4}{2}},2^{\frac{n-4}{2}}), p, q\neq 0;\\ \U(\frac{n-2}{2}), (n,0), (0,n). \end{array}$
         & $\GL(2^{\frac{n-4}{2}},\H)$    \\ \hline
\end{tabular}
\end{center}

\medskip

Note that if we know isomorphisms between these Lie groups, then we know isomorphisms between corresponding Lie algebras.
So, we also obtain isomorphisms between Lie algebras $\overline{\textbf{2}}\oplus i\overline{\textbf{1}}$, $\overline{\textbf{2}}\oplus i\overline{\textbf{3}}$, $ \overline{\textbf{2}}\oplus\overline{\textbf{3}}$, $\overline{\textbf{2}}\oplus\overline{\textbf{1}}$, $\overline{\textbf{2}}$ and corresponding classical matrix Lie algebras: linear $\gl(k,\R)$, $\gl(k,\H)$, unitary $\u(k)$, $\u(r,s)$, orthogonal $\so(k,\R)$, $\so(r,s)$, $\so(k,\C)$, $\so(k,\H)$, symplectic $\sp(k)$, $\sp(r,s)$, $\sp(k,\R)$, $\sp(k,\C)$ or direct sums of such Lie algebras of corresponding dimensions in different cases.

As we have already mentioned in the Introduction, group $G^\epsilon_{p,q}$ \cite{Abl2} coincides with the group $\G^{23}_{p,q}$ in the case of signature $(n,0$) and with the group $\G^{12}_{p,q}$ in the case of signature $(0,n)$. In \cite{Lounesto} (p. 236, Tables 1 and 2) one finds isomorphisms between classical matrix Lie groups and the groups $\G^{12}_{p,q}$ and $\G^{23}_{p,q}$ which are automorphism groups of the scalar products on the spinor spaces. In the present paper we use another technique. We obtain the same isomorphisms for groups $\G^{12}_{p,q}$ and $\G^{23}_{p,q}$ and also isomorphisms for groups $\G^{2i1}_{p,q}$, $\G^{2i3}_{p,q}$, $\G^{2}_{p,q}$ using the notion of additional signature and relations between operations of conjugation in Clifford algebra and the corresponding matrix operations. We also study the corresponding Lie algebras which are related to subspaces of quaternion types.

\end{document}